\documentclass[12pt,preprint2]{aastex}
\usepackage{natbib}
\usepackage{color}

\newcommand{\hr}{HR~8799 }

\newcommand{\hrb}{HR~8799b }

\newcommand{\hrbnosp}{HR~8799b}

\shorttitle{\hr in M-band}
\shortauthors{Galicher et al.}

\begin{document}

\title{M-band Imaging of the \hr Planetary System Using an Innovative LOCI-based Background Subtraction Technique}

\author{Rapha\"el Galicher\altaffilmark{1,2}, Christian Marois\altaffilmark{1}, Bruce~Macintosh\altaffilmark{3}, Travis Barman\altaffilmark{4} and Quinn Konopacky\altaffilmark{3}}
\affil{\altaffilmark{1}National Research Council Canada, Herzberg Institute of Astrophysics, 5071 West Saanich Road, Victoria, BC, V9E 2E7, Canada;}
\affil{\altaffilmark{2}Dept. de Physique, Universit\'e de Montr\'eal, C.P. 6128 Succ. Centre-ville, Montr\'eal, Qc, H3C 3J7, Canada;}
\affil{\altaffilmark{3}Lawrence Livermore National Laboratory, 7000 East Ave., Livermore, California, 94550, USA;}
\affil{\altaffilmark{4}Lowell Observatory, 1400 West Mars Hill Road, Flagstaff, AZ 86001, USA;}
    \email{raphael.galicher@nrc-cnrc.gc.ca}

\begin{abstract}
Multi-wavelength observations/spectroscopy of exoplanetary atmospheres are the basis of the emerging exciting field of comparative exoplanetology. The \hr planetary system is an ideal laboratory to study our current knowledge gap between massive field brown dwarfs and the cold 5-Gyr old Solar system planets. The \hr planets have so far been imaged at J- to L-band, with only upper limits available at M-band. We present here deep high-contrast Keck II adaptive optics M-band observations that show the imaging detection of~$3$ of the~$4$ currently known \hr planets. Such detections were made possible due to the development of an innovative LOCI-based background subtraction scheme that is~$3$ times more efficient than a classical median background subtraction for Keck II AO data, representing a gain in telescope time of up to a factor of~$9$. These M-band detections extend the broad band photometric coverage out to~$\sim 5\mu$m and provide access to the strong CO fundamental absorption band at~$4.5\mu$m. The new M-band photometry shows that the \hr planets are located near the L/T-type dwarf transition, similar to what was found by other studies. We also confirm that the best atmospheric fits are consistent with low surface gravity, dusty and non-equilibrium CO/CH$_4$ chemistry models.
\end{abstract}
    
    \date{ }
        
    \maketitle
    
\section{Introduction}
After more than a decade of searching, the direct exoplanet imaging quest was finally successful in 2008 with the discovery of three planetary systems \citep{marois08,kalas08,lagrange09}. One key advantage of this technique is the detection of the planet's thermal emission. Detailed multi-band photometry and spectrometry can be acquired that can then be compared with atmospheric models to derive the planet's physical characteristics and study the effect of dust and molecular chemistry. With its multiple co-eval Jovian planets, the \hr system is an ideal laboratory to study young planets with low temperature/surface gravity atmospheres and bridge the gap between massive field brown dwarfs and the cold Solar system planets.

\hr is a 30Myr old~\citep{marois10,zuckerman11} A5V star located 39.4pc away \citep{leeuwen07} in the Pegasus constellation. It is classified as a~$\gamma$ Doradus and a $\lambda$ Bootis star~\citep{gray99}. It also shows an IR excess~(Vega-like star) consistent with a debris disk composed of a warm dust disk~(6 to 15AU), a massive cold dust disk~(90 up to 300AU) and a small dust particle halo extending up to a 1,000AU \citep{rhee07,su09}.

Photometry and some spectroscopy of the \hr planets are available in several bands from 1 to 3.8$\mu$m~\citep{marois08, lafreniere09,metchev09,janson10,marois10, hinz10,currie11,barman11}. An initial characterization \citep{marois08,bowler10,marois10,currie11,barman11} has been achieved using the available data and state-of-the-art atmospheric models developed for field brown dwarf analysis. It is clear that these planets are quite different than most field brown dwarfs, showing very dusty atmospheres while being cool ($\sim $1,000K) with no sign of methane molecular absorption.  L- and M-band detections/upper limits suggest this lack of methane is due to a CO/CH$_4$ non-equilibrium chemistry \citep{saumon03,cushing06,leggett07,hinz10,bowler10,currie11,barman11}. Other teams have been able to fit the available photometry with patchy clouds and equilibrium chemistry atmospheres ~ \citep{marois08,currie11}. Accurate M-band photometry can help disentangle the various models, but it is hard on a ground-based telescope due to the bright thermal background mainly from the telescope and the adaptive optics system~\citep{Lloyd00}. This background is particularly hard to remove at Keck, as it varies with time as the instrument image rotator (which is located very near focus) moves and as the AO deformable mirror modulates the rotator-induced background pattern as seen by the detector. As a result, conventional sky/background subtraction routines leaves a spatially variable residual that limits final sensitivity.

In this letter, we present the first M-band ($4.670 \mu$m) detections of three of the currently four known \hr planets, the longest wavelength at which these planets have been imaged. The observations are discussed in~\S\ref{sec : obs}. The data reduction along with the new very efficient LOCI-based (LOCI: locally optimized combination of images) background subtraction technique are presented in~\S\ref{sec : reduc}. Comparisons with field brown dwarfs and new atmospheric fits are shown in \S\ref{sec : dis}. Conclusions are summarized in~\S\ref{sec : conc}.
        
\section{Observations}
\label{sec : obs}
The \hr data was obtained on 2009 Nov.~1 and 2 at the Keck II observatory using the adaptive optics system, the NIRC2 near-infrared narrow field camera~\citep{mclean03} and the M-band filter~($\lambda_0=4.670\,\mu$m, $\Delta\lambda=0.241\,\mu$m). No coronagraph was used for these observations. For these two nights, 140~unsaturated images~(58 for Nov. 1 and~82 for Nov. 2) were acquired, each image consisting of 200 coadds of 0.3s, for a total on-source integration time of 140min. The data was taken in angular differential imaging mode~\citep{marois06}. In that mode, the instrument field rotator is adjusted to track the telescope pupil while leaving the field-of-view~(FOV) to rotate with time. In addition, the telescope was nodded every few minutes between four positions to facilitate the background subtraction. In our data, the total FOV rotation is~$\sim 180$\,degrees for both nights, but both data sets were not acquired through transit (between HA [-1.6h,-0.4h] and [0.6h,1.1h] for Nov.~1 and [-1h,-0.6h] and [0.5h,2.2h] for Nov.~2). Although such observing strategy is fine for background limited data, it did limit the amount of speckle subtraction achievable at smaller separations where the data is still speckle noise limited and where planet e is now known to be located ( the \hr transit time was actually used during these two nights for L-band imaging to confirm planet e).

\section{Data reduction}
    \label{sec : reduc}
The data reduction was performed as follows: we first subtract a dark image, divide by a flat field and mask bad/hot pixels. We then remove known NIRC2 narrow-field camera distortions using the \citet{yelda10} solution. The images are rotated by 0.252 degrees clockwise to put North up and the plate scale is selected to be $9.952 \pm 0002$mas/pixel to be consistent with the distortion correction of \citet{yelda10}. After these steps, we subtract the background~(\S\,\ref{subsec : background}) and the speckle noise~(\S\,\ref{subsec : spec}) on the two data cubes of 58~(Nov.~1) and 82 (Nov.~2) images separately. We then average the two processed sets to produce a final image.

\subsection{LOCI-based background subtraction}
\label{subsec : background}
\begin{figure}[!ht]
\centering
\includegraphics{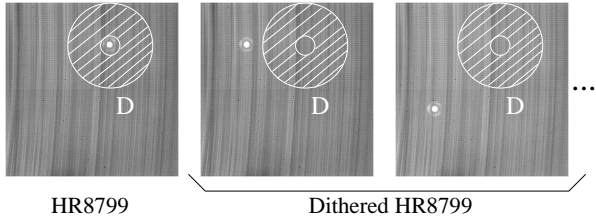}
\caption{\sl Image~$I_p$ where the background has to be subtracted~(left) and the~D region inside which the LOCI coefficients are optimized using the reference background dither images ~(right).}
\label{fig : f1}
\end{figure}

The telescope was nodded between exposures to record the star at different detector locations. For each image of the sequence, a classical background reduction would consist of subtracting the median of all the images where the star is sufficiently dithered. This standard technique is limited because of the time variations of the thermal background whereas the median uses all the reference background images without distinction. As the background subtraction step is no different than a standard ADI/LOCI speckle subtraction, we have developed a background subtraction routine~\citep{marois10b} based on the locally optimized combination of images algorithm~\citep[LOCI,][]{lafreniere07}. This new routine is more efficient than the classical median subtraction as it puts different weights on the background reference images. Consider the background subtraction in the~$p$th image~$I_p$ of the sequence. An annulus~(called~$D$, Fig.~\ref{fig : f1}) is defined around the star and all reference background images are selected where the star has been sufficiently dithered compared to $I_p$.
The algorithm searches for the linear combination of these images that minimizes~the $\chi^2_p$~(i.e. the residual noise in~$D$ excluding bad pixels) which can be expressed as
\begin{equation}
\chi_p^2(\alpha^{1\le i\le N}_p) = \sum_{i=1}^{N}\sum_{j=1}^{M}[\alpha^i_p\,I_i(x_j)-I_p(x_j)]^2\,D(x_j)
\end{equation}
where~$M$ is the number of pixels in the image,~$N$ is the total number of images in the sequence, $x_j$ is the $j$th pixel, $I_i$ is the~$i$th image of the sequence, and~$\alpha^i_p$ the coefficients to be determined. The inner radius of area~$D$ is chosen to avoid the bright star speckle noise that can bias the background minimization algorithm, while its outer radius is chosen to avoid the dithered star images. In our case, the inner and outer radii are 3 and 30\,$\lambda/$D corresponding to 24 to 240pixels. We have then imposed~$\alpha_p^i=0$ if the star is at a similar location~in~$I_i$ as in~$I_p$. The LOCI algorithm then finds the optimal~$\alpha_p^{i,best}$~\citep{lafreniere07,marois10b} that minimizes~the~$\chi_p^2$.  The optimized background image~$B_p$ at each pixel for~$I_p$~(not only in~$D$) is then
\begin{equation}
B_p=\sum_{i=1}^{N}\alpha_p^{i,best}\,I_i.
\end{equation}
The generated reference background image $B_p$ is finally subtracted from~$I_p$. As~$\alpha_p^{i,best}=0$ for the images where the star is registered at the same location in~$I_i$ as in~$I_p$, $B_p$ does not contain any flux from a possible companion. Moreover, if a companion exists in~$I_p$, its impact on the~$\alpha_p^{i,best}$ LOCI coefficients is negligible since the companion flux is much smaller than the thermal background and~$D$ is large compared to the companion~PSF.

Once the background is subtracted, we sub-pixel register each image using an iterative cross-correlation gaussian fit to the PSF core~(images are unsaturated).

\subsection{Median speckle subtraction}
\label{subsec : spec}
\begin{figure*}[!ht]
\centering
\includegraphics{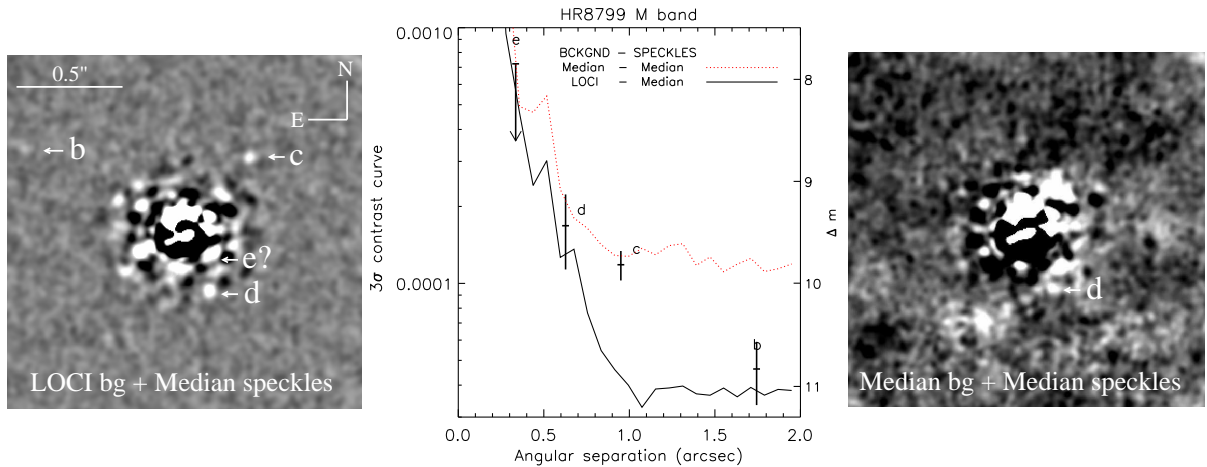}
\caption[]{\sl Left panel: Final image (average of the Nov. 1 and 2 data) after subtracting the background noise with our new LOCI-based algorithm and the speckle noise with a basic median. Right panel: Same reduction, but using a classical median background subtraction. We applied an unsharp mask~(median in a~$4\times4 \lambda/$D box) on the two images and then convolved them by a~$0.5\lambda/D$ width Gaussian. The two panels have the same linear intensity scale and~FOV. North is up and East is left. Central panel: $3 \sigma$ radial contrast noise profiles of our resulting median-combined images after subtracting the background with LOCI processing~(black full line) or a median~(red dashed line). Planets b, c and d fluxes are also plotted with $1 \sigma$ error bars along with the planet e flux upper limit.}
\label{fig : f2}
\end{figure*}

The speckle subtraction is performed following a basic~ADI data reduction as described in~\citet{marois06}. For each data cube, an initial reference PSF is assembled by taking the median of all images and is then subtracted. We then rotate the images to put North up and median combine them. We apply an unsharp mask to the resulting image~(median in a~$4\times4 \lambda/$D box) to remove the low spatial frequency noise and convolve it by a~$0.5 \lambda$/D width Gaussian to average the high frequency pixel-to-pixel noise. We finally average the Nov.~1 and 2 images (Fig.~\ref{fig : f2}).
	
Without the~LOCI background subtraction, none of the planets would have been detected (right panel, signal-to-noise ratio (SNR) for~d is less than~$2$). With the~LOCI background subtraction, \hrbnosp, c and d are detected~(left panel, $3$ to $8$~SNR). Planet e non-detection is probably due to both sequences not being acquired through transit, thus limiting the amount of speckle noise being removed at small separations from the median subtraction. We tried to apply a more advanced LOCI algorithm~\citep{lafreniere07,marois08b,marois10b} to improve the speckle reduction, but as the~FOV rotation ranges were small for both nights, no contrast gain was achieved.
 
\section{Data analysis}
\label{sec : res}
Planet fluxes and positions were obtained by subtracting the planets prior to the speckle reduction using the stellar unsaturated PSF as the template. We also tried subtracting the companions prior to the LOCI-background algorithm and we have confirmed that no bias is introduced by this technique~(final flux variations smaller than~$0.07\%$). The subtraction was iterated by moving the planet template and changing its intensity until a minimal noise residual at the planet's location was achieved (inside a 1.5~$\lambda$/D radius area; Tab.~\ref{tab : tab1} for the resulting magnitudes). Photometric error bars were calculated in~$\lambda/D$ width annulus. As expected from other wavelengths~\citep{marois08}, \hrb flux is roughly a third of that of planets~c and~d. The planet's positions are included in Tab.~\ref{tab : tab1}. The low SNRs and the large M-band PSF core result in large astrometric errors. A future astrometric HR~8799 paper using shorter wavelength astrometry is in preparation.

\begin{table*}[!ht]
\begin{center}
\begin{tabular}{lcccc}
\tableline
\tableline
Planet &{b}&c&d&e\\
\tableline
Separation w.r.t HR8799a (E,N)"&$(1.54,0.80)$&$(-0.63,0.72)$&$(-0.24,-0.58)$&\textsuperscript{b}\\
&$\pm0.019$&$\pm0.013$&$\pm0.014$&\\
Contrast $\Delta\,m$&$10.84\pm0.30$&$9.82\pm0.14$&$9.44\pm0.35$&$>7.85$\textsuperscript{a}\\
Absolute $M$&$13.07\pm0.30$&$12.05\pm0.14$&$11.67\pm0.35$&$>10.09$\textsuperscript{a}\\
Absolute $M$~\citep{currie11}&$>11.37$\textsuperscript{a}&$>11.22$\textsuperscript{a}&$>11.15$\textsuperscript{a}&\\
Absolute flux (mJy)&$0.91\pm0.21$&$2.33\pm0.25$&$3.31\pm0.65$&$<14.23$\textsuperscript{a}\\ Currie's absolute flux (mJy)&$<4.36\textsuperscript{a}$&$<5.00\textsuperscript{a}$&$<5.34\textsuperscript{a}$&\\
\end{tabular}
\caption{\sl \hr planets M-photometry. HR~8799a apparent magnitude is~$5.21$ and its distance is~$39.4\pm1.0$pc. Magnitude zeropoint is~$154$\,Jy.~\citep{cox00}. \textsuperscript{a}~$3\sigma$ upper limits. \textsuperscript{b}~We use the \citet{marois10} astrometry for planet~e.}   \label{tab : tab1}
\end{center}
\end{table*}

Contrast plots~(central panel of~Fig.~\ref{fig : f2}) were obtained by calculating the noise in an annulus having a $\lambda/$D width normalized by the stellar PSF flux (after performing the same unsharp mask and convolution of a 0.5 $\lambda/$D Gaussian). The  contrast plots were then normalized at each separation by the estimated point source throughput using simulated ADI median process planets. Fig.~\ref{fig : f2} shows that the LOCI background subtraction~(black full line) is up to~$\sim 3$ times better than a classical background median subtraction~(red dotted line). If a classical background subtraction routine is used, an integration time of up to 9 times longer is required to reach the same LOCI background subtraction contrast. This new highly efficient LOCI-based background subtraction routine can be used on any data where the background is non-negligible and evolving with time.

\section{Discussions}
\label{sec : dis}
\begin{figure}[!ht]
\centering
\includegraphics{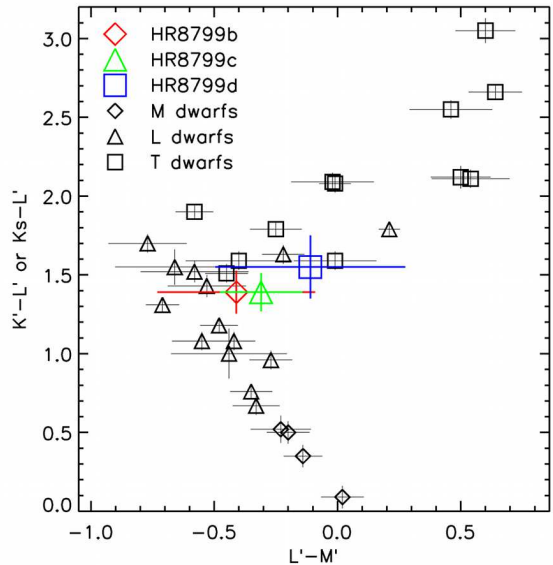}
\caption[]{\sl K-L vs L-M diagram for field brown dwarfs, lower mass stars and the \hr planets. The planets are located near the end of the L-type sequence, close to the T-type transition.}
\label{fig : f3}
\end{figure}

It has been shown that the \hr planets are an L-type extension towards lower effective temperatures and lower surface gravities~\citep{marois08,bowler10,currie11,barman11}. The planets have also been found to be dusty with evidence of non-equilibrium CO/CH$_4$ chemistry. If the \hr planets are plotted in  K'-L' vs L'-M' diagram~(Fig.~\ref{fig : f3}) against field brown dwarfs and lower mass stars~\citep{leggett07}, it is found that the \hr planets are located near the L/T-type dwarf transition. The M-band photometry is thus consistent with what was found previously by other studies.
From the field brown dwarf best fit parameters of~Fig.~4-6 of \citet{leggett07}, we also find that the planet L- and M-band colors are consistent with dusty atmospheres with vertical mixing. \citet{barman11} found a similar result for~\hrb while considering lower surface gravities.

If we compare our estimated M-band flux to~\citet{currie11} equilibrium chemistry model predictions, we find that all their best patchy-cloud atmospheric fits are rejected to~$\sim 3\sigma$ for~\hrbnosp. For planets~c and~d, \citet{currie11} use equilibrium chemical models and focus on clouds to interpret the available photometry. The earlier work of~\citet{hinz10} did compare different chemical models to their M-band upper limits for all three planets, but did not include near-IR photometry. Our new photometry is consistent with some models proposed by the two teams and will help in constraining their fits.

\begin{figure*}[!ht]
\centering
\includegraphics{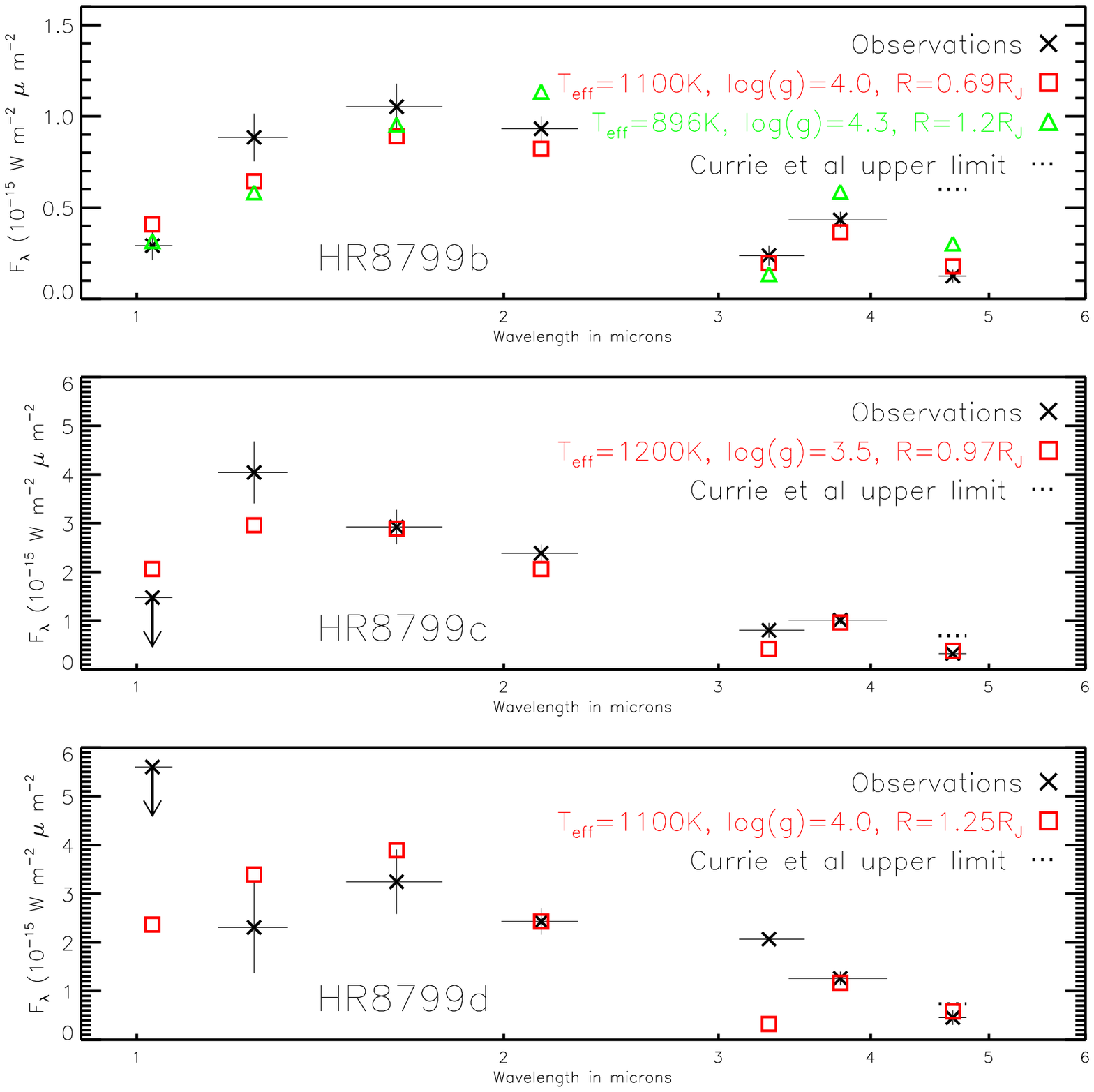}
\caption[]{\sl Comparison between the best solar-abundance model~(red squares) and the broad band photometry~(black cross) for \hrb(top), c~(middle) and~d~(bottom). For planet~b, the photometry from the~$10$ times more metal rich case described in~\citet{barman11} is plotted in green triangles. The photometric measurements other than for the M-band are from~\citet{marois08,marois10,currie11}. We also show~\citet{currie11} upper limits for the M-band photometry as dotted-line.}
\label{fig : f4}
\end{figure*}

In this letter, we compare cloudy and non-equilibrium model fits for all of the photometry for the three planets.  Such comparison has only been done for planet~b so far~\citep{bowler10,barman11}. Using the same solar abundance model atmosphere grids and procedures described in~\citet{barman11},  new gravities and effective temperatures are found (Fig.~\ref{fig : f4}).
For planet~b, we find similar model atmosphere parameters as in~\citet{barman11}, but with a slightly higher gravity -- $log(g)=4$ rather than~$3.5$.  This solar abundance model has $T_{\mathrm{eff}}=1100K$ and, thus, requires a very small radius ($\sim0.7\,R_j$) to match the observed bolometric luminosity; nearly a factor of~$2$ smaller than predicted by traditional hot-start models~\citep{baraffe03}.
For planets~c and~d, we find temperatures and gravities that are fairly close to the expected values from evolution models~\citep{marois08}. \citet{barman11} argue that increasing the metallicity to~$10$ times solar could bring the $T_{\mathrm{eff}}/log(g)$ derived from atmosphere and evolution models for planet~b into better agreement. The photometry from this metal rich, evolution-consistent, model is also shown in Fig.~\ref{fig : f4}.  This model, however, predicts an M-band flux that is~$3.5\sigma$ brighter than observed~(while all other bands agree to~$2\sigma$ or better). This may indicate an even higher overall metallicity and/or that a non-solar~C-to-O ratio is required.  A broader exploration of the possible metal abundances for all four planets will be left to a future paper.

 \section{Conclusions}
 \label{sec : conc}
In this letter, we have estimated for the first time the M-band fluxes of three of the currently four known \hr planets. These detections were made possible due to the use of an innovative LOCI-based background subtraction routine that has allowed for a factor of~$3$ gain in contrast (factor of~$9$ in integration time) compared to a classical background subtraction using a median. This new background subtraction routine can be used to subtract the background noise in any infrared data.

We have detected \hrbnosp, c and~d at M-band from 3 to 8$\sigma$. From a K'-L' and L'-M' color diagram, we confirm that the three planets are located near the end of the L-type sequence, close to the L- and T-dwarf transition region. We then derived new atmosphere model fits for the three planets. For planets~c and~d, temperatures and surface gravities are close to the expected values from evolutionary models. For planet~b, the solar abundance model fits well the broad band photometry from~$1$ to~$5\,\mu$m but it requires a very small planetary radius to match the bolometric luminosity and is thus in contradiction with the planet evolution models. The metal rich evolution-consistent model over-predicts the M-band flux, which may indicate an even higher overall metallicity and/or a non-solar C-to-O ratio. Higher~SNR images will help disentangle the different physical and chemical parameters.
      
\section{Acknowledgment}      
The authors wish to thank S. Leggett for kindly providing the~Fig.~\ref{fig : f3} field brown dwarf data. Portions of this work performed under the auspices of the U.S. Department of Energy by Lawrence Livermore National Laboratory under Contract DE-AC52-07NA27344. The data presented herein were obtained at the W.M. Keck Observatory, which is operated as a scientific partnership among the California Institute of Technology, the University of California and the National Aeronautics and Space Administration. The Observatory was made possible by the generous financial support of the W.M. Keck Foundation. The authors wish to recognize and acknowledge the very significant cultural role and reverence that the summit of Mauna Kea has always had within the indigenous Hawaiian community.  We are most fortunate to have the opportunity to conduct observations from this mountain.


\begin{thebibliography}{...}
    
    \bibitem[Baraffe et al.(2003)]{baraffe03}
        Baraffe, I., et al., 2003, {\sl Astronomy and Astrophysics}, {\bf 402}, 701--712.
    
    \bibitem[Barman et al.(2011)]{barman11}
    	Barman, T., et al., 2011, {\sl The Astrophysical Journal}, {\bf 733}, 65--.
	
    \bibitem[Bowler et al.(2010)]{bowler10}
    	Bowler, B.~P., et al., 2010, {\sl The Astrophysical Journal}, {\bf 723}, 850--868.
    
     \bibitem[Cox(2000)]{cox00}
        Cox, A.~N., 2000, {\sl Allen's astrophysical quantities}.

     \bibitem[Cushing et al.(2006)]{cushing06}
        Cushing, M.~C., et al., 2006, {\sl The Astrophysical Journal}, {\bf 648}, 614-628.
    
    \bibitem[Currie et al.(2011)]{currie11}
    	Currie, T., et al., 2011, {\sl submitted}.

    \bibitem[Gray et al.(1999)]{gray99}		
	Gray, R.~O., et al., 1999, {\sl The Astronomical Journal},{\bf 118},2993--2996
	
    \bibitem[Hinz et al.(2010)]{hinz10}
    	Hinz, P.~M., et al., 2010, {\sl The Astrophysical Journal}, {\bf 716}, 417--426.

    \bibitem[Kalas et al.(2008)]{kalas08}
    	Kalas, P., et al., 2008, {\sl Science}, {\bf 322}, 1345--.

    \bibitem[Janson et al.(2010)]{janson10}
        Janson, M., et al., 2010, {\sl The Astrophysical Journal Letters}, {\bf 710}, L35--L38.

     \bibitem[Lafreni\`ere et al.(2007)]{lafreniere07}
       Lafreni\`ere, D., et al., 2007, {\sl The Astrophysical Journal}, {\bf 660}, 770--780.
      
     \bibitem[Lafreni\`ere et al.(2009)]{lafreniere09}
       Lafreni\`ere, D., et al., 2009, {\sl The Astrophysical Journal}, {\bf 694}, L148--152.

    \bibitem[Lagrange et al.(2009)]{lagrange09}
    	Lagrange, A.~M., et al., 2009, {\sl Astronomy and Astrophysics}, {\bf 493}, L21-L25.

     \bibitem[Leeuwen et al.(2007)]{leeuwen07}
     	Leeuwen, E. Van, 2007, {\sl Astronomy and Astrophysics}, {\bf 474}, 653--.

      \bibitem[Leggett et al.(2007)]{leggett07}
        Leggett, S., et al., 2007, {\sl The Astrophysical Journal}, {\bf 655}, 1079-1094.

        \bibitem[Lloyd(2000)]{Lloyd00}
          Lloyd-Hart, M., 2000, {\sl The Publications of the Astronomical Society of the Pacific}, {\bf 112}, 264--272.

    \bibitem[Marois et al.(2006)]{marois06}
      Marois, C., et al., 2006, {\sl The Astrophysical Journal}, {\bf 641}, 556--564.
    
    \bibitem[Marois et al.(2008)]{marois08}
      Marois, C., et al., 2008, {\sl Science}, {\bf 322}, 1348--.

    \bibitem[Marois et al.(2008b)]{marois08b}
     Marois, C., et al., 2008b, {\sl The Astrophysical Journal}, {\bf 673}, 647--656.

    \bibitem[Marois et al.(2010)]{marois10}
      Marois, C., et al., 2010, {\sl Nature}, {\bf 468}, 1080--1083.

    \bibitem[Marois et al.(2010b)]{marois10b}
      Marois, C., et al., 2010b, {\sl SPIE}, {\bf 7736}.
   
    \bibitem[McLean et al.(2003)]{mclean03}
      McLean, I.~S., and Sprayberry, D., 2003, {\sl SPIE}, {\bf 4841}.

      \bibitem[Metchev et al.(2009)]{metchev09}
      Metchev, S., et al., 2009, {\sl The Astrophysical Journal Letters}, {\bf 705}, L204--L207.
      
    \bibitem[Rhee et al.(2007)]{rhee07}
	Rhee, J.~H., et al., 2007, {The Astrophysical Journal}, {\bf 660}, 1551--1571.
	           
     \bibitem[Saumon et al.(2003)]{saumon03}
        Saumon, D., et al., 2003, {\sl IAU Symposium}, {\bf 211}, 345--.

     \bibitem[Su et al.(2009)]{su09}
        Su, K.~.Y.~L., et al., 2009, {\sl The Astrophysical Journal}, {\bf 705}, 314--327.
       
     \bibitem[Yelda et al.(2010)]{yelda10}
        Yelda, S., et al., 2010, {\sl The Astrophysical Journal}, {\bf 725}, 331--352.
      
      \bibitem[Zuckerman et al.(2011)]{zuckerman11}
        Zuckerman, B., et al., 2011, {\sl The Astrophysical Journal}, {\bf 732}, 61--+.
         
    \end{thebibliography}
\end{document}